\documentclass[aip,jcp,reprint,floatfix]{revtex4-1}
\usepackage{amsfonts,amssymb,amsmath,array}
\usepackage{dcolumn}
\usepackage{bm}
\usepackage[final]{graphicx}
\usepackage{graphicx}
\usepackage{epstopdf}
\graphicspath{{./Figures/}} 

\usepackage[usenames]{color} 

\usepackage{appendix}

\begin{document}

\title{The Parental Active Model: a unifying stochastic description of self-propulsion}

\author{Lorenzo Caprini}
\email{lorenzo.caprini@gssi.it, lorenzo.caprini@hhu.de}
\affiliation{Institut f\"ur Theoretische Physik II: Weiche Materie, Heinrich-Heine-Universit\"at D\"usseldorf, 40225 D\"usseldorf, Germany}

\author{Alexander R. Sprenger}
\affiliation{Institut f\"ur Theoretische Physik II: Weiche Materie, Heinrich-Heine-Universit\"at D\"usseldorf, 40225 D\"usseldorf, Germany}

\author{Hartmut L{\"o}wen}
\email{hlowen@hhu.de}
\affiliation{Institut f\"ur Theoretische Physik II: Weiche Materie, Heinrich-Heine-Universit\"at D\"usseldorf, 40225 D\"usseldorf, Germany}

\author{Ren{\'e} Wittmann}
\email{rene.wittmann@hhu.de}
\affiliation{Institut f\"ur Theoretische Physik II: Weiche Materie, Heinrich-Heine-Universit\"at D\"usseldorf, 40225 D\"usseldorf, Germany}

\begin{abstract}
We propose a new overarching model for self-propelled particles that flexibly generates a full family of ``descendants''. 
The general dynamics introduced in this paper, which we denote as ``parental'' active model (PAM), unifies two special cases commonly used to describe active matter, namely active Brownian particles (ABPs) and active Ornstein-Uhlenbeck particles (AOUPs). 
We thereby document the existence of a deep and close stochastic relationship between them, resulting in the subtle balance between fluctuations in the magnitude and direction of the self-propulsion velocity. 
 Besides illustrating the relation between these two common models, the PAM 
can generate additional offspring, interpolating between ABP and AOUP dynamics, that could provide more suitable models for a large class of living and inanimate active matter systems, 
possessing characteristic distributions of their self-propulsion velocity.
Our general model is evaluated in the presence of a harmonic external confinement. 
For this reference example, we present a two-state phase diagram which sheds light on the transition in the shape of the positional density distribution, from a unimodal Gaussian for AOUPs to a Mexican-hat-like profile for ABPs.
\end{abstract}

\maketitle

\paragraph*{Introduction.}
Active matter includes a broad variety of biological and physical systems~\cite{bechinger2016active, marchetti2013hydrodynamics,  elgeti2015physics}, ranging from bacteria~\cite{arlt2018painting, frangipane2018dynamic}, colloids~\cite{yan2016,ni2017,driscoll2017,ginot2018,khadka2018,stoop2018}, more complex organisms such as sperms and cells~\cite{alert2020physical}, and even animals at the macroscopic scales~\cite{couzin2002collective, zampetaki2021collective} such as birds~\cite{cavagna2018physics} and fish~\cite{perna2014duality}. 
Each of these systems is formed by individual active units which convert energy into motion, a property which allows them to be denoted as active systems~\cite{gompper20202020}. 
Despite this generic label, the multitude of mechanisms behind active motion results in a large amount of diversity, e.g., giving rise to systems whose typical active velocity is constant or subject to fluctuations.

On the theory side, there are two major paradigms for modeling active particles as a diffusive stochastic process~\cite{fodor2018statistical}: active Brownian particles (ABPs)~\cite{buttinoni2013dynamical, fily2012athermal, stenhammar2014phase, bialke2015negative, solon2015pressure, petrelli2018active, caprini2020hidden}, introduced to describe the diffusion-driven behavior of active colloids, and active Ornstein-Uhlenbeck particles (AOUPs)~\cite{ maggi2015multidimensional,caprini2018activeescape,dabelow2019irreversibility,berthier2019glassy, wittmann2018effective, fily2019, mandal2017entropy, fodor2016far, martin2021statistical}, proposed for mathematical convenience but also found to be a good approximation for a passive particle in an active bath~\cite{maggi2014generalized, maggi2017memory, chaki2019effects, goswami2021work}.
Both models posses two major common ingredients: the typical self-propulsion velocity induced by the active force (sometimes called swim velocity), which is constant for ABPs or given by an average value for AOUPs, and the persistence time, indicating the strength of rotational diffusion for ABPs and the characteristic time-scale in the autocorrelation of the active noise for AOUPs.

\begin{figure}[t!]
\includegraphics[width=\columnwidth]{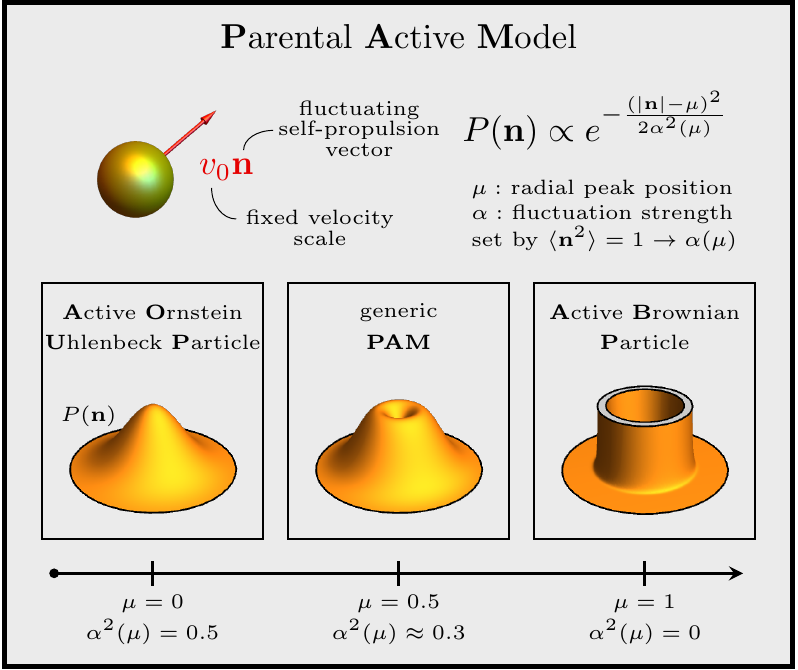}
\caption{Illustration of the considered family of active models, uniquely characterized by a velocity scale $v_0$ and the self propulsion vector $\mathbf{n}$, determined by a stochastic process of unit variance. 
The parental active model (PAM) is described by the shown distribution $P(\mathbf{n})$ in the form of a shifted Gaussian (see Eq.~\eqref{eq_PAM}) with the single free parameter $\mu$, 
which identifies the most likely value of the modulus $|\mathbf{n}|$. 
The width of the distribution, quantified by $\alpha(\mu)$, is constrained by the condition $\langle \mathbf{n}^2\rangle=1$ (see Eq.~\eqref{eq:alphamu_approx}).  
The 3d plots at the bottom show $P(\mathbf{n})$ 
for three specific choices of $\mu$, indicated by the axis below, which are further discussed in the text. 
}
\label{fig:picture}
\end{figure}

It is well known that ABPs and AOUPs share a similar phenomenology in a large range of fundamental physical problems, e.g., both predict the accumulation near walls and obstacles~\cite{caprini2018active, das2018confined, caprini2019activechiral}, clustering \cite{palacci2013living,mognetti2013living} and motility induced phase separation~\cite{buttinoni2013dynamical, solon2015pressure, paliwal2018,van2019interrupted, shi2020self, turci2021phase, caprini2020spontaneous, maggi2021universality}, as well as spatial velocity correlations in dense systems~\cite{caprini2020hidden, szamel2021long, caprini2020time, caprini2021spatial}.
However, some prominent differences emerge in a few special cases, such as the failure of AOUPs to reproduce the bimodal spatial distribution in a harmonic potential (for instance, see Ref.~\onlinecite{szamel2014self} for AOUPs and Refs.~\onlinecite{takatori2016acoustic, malakar2020steady} for ABPs), or the distinct behavior of the density in the bulk of a confined system~\cite{yan2015force, fily2017equilibrium, wittmann2019pressure}.
For this reason, ABPs are usually perceived as the established model to describe active colloids, while AOUPs are considered as a useful but oversimplified approximation for ABPs  
when the model parameters are appropriately chosen. 
 However, the propitious theoretical possibilities offered by the AOUPs contributed to establish it as an important model for active matter systems in its own right.
 This has lead to a continuously increasing number of works dedicated to the AOUP model, with the aim of deriving exact or approximate analytical results for single-particle~\cite{caprini2020inertial, nguyen2021active}
or interacting systems~\cite{farage2015effective,marconi2016pressure, marconi2016velocity, wittmann2016active,wittmann2017effective}.
The recent interest in AOUPs implies the need to reevaluate the unilateral relation to the ABP model by going beyond the standard qualitative way to compare these two fundamental approaches.

 In this work, we  propose a general model to describe the self-propulsion mechanism of active particles  on the microscale, which we call parental active model (PAM) because it includes both ABPs and AOUPs as two subcases.
We thus show that these classical models actually stand on the same hierarchical level as descendants of the PAM, see Fig.~\ref{fig:picture} for an illustrative picture.
Specifically, they differ only by the value of a single parameter, indicating the shape of the probability distribution of the radial component of the active velocity.
In other words, the relation between ABPs and AOUPs is that of two sisters rather than two cousins.
By considering a whole class of overarching models, we both uncover the deep connection between ABPs and AOUPs, going beyond a mutual mapping~\cite{farage2015effective, caprini2019comparative}, and bridge the gap between these two extreme cases, which may provide a crucial step towards a more realistic description of experimental systems.
To explore the whole family of models, we compare the (famously distinct) probability density of ABPs and AOUPs in a harmonic trap to the results for intermediate offspring of the PAM.

\paragraph*{Generic dynamics of active particles.} 

The typical overdamped dynamics of a generic active particle is described by the following differential equation for its position $\mathbf{x}$:
\begin{equation}
\gamma\dot{\mathbf{x}}= \gamma v_0\mathbf{n} + {\gamma}\sqrt{2 D_t} \mathbf{w} + \mathbf{F}(\mathbf{x}) \,,
\label{eq_t1}
\end{equation}
where $\mathbf{F}(\mathbf{x})$ is the external force exerted on the particle, $\mathbf{w}$ is a white noise with unit variance and zero average.
$\gamma$ and $D_t$ are the friction coefficient and the translational diffusion coefficient, respectively, related to the temperature of the bath through the Einstein relation.
The term $v_0 \gamma \mathbf{n}$ is called active force and $v_0 \mathbf{n}$ is the resulting self-propulsion velocity, where the constant $v_0$ provides a velocity scale. 
The self-propulsion vector $\mathbf{n}$ is a general stochastic process with unit variance, whose specific dynamics determine the active model under consideration.
For simplicity we restrict ourselves to two spatial dimensions.

\paragraph*{Active Brownian Particles (ABPs).} In the case of ABPs, $\mathbf{n}$ represents a unit vector which denotes the fluctuating particle orientation. 
In other words, the direction of $\mathbf{n}=(\cos\theta,\sin\theta)$ is described by the following steady-state distribution: 
\begin{equation}
P_\text{ABP}(n, \theta) \sim \frac{1}{2\pi} n \delta(n - 1) 
\end{equation}
with a uniformly distributed orientational angle $\theta$ and fluctuation-free modulus $n=|\mathbf{n}|$ that is  always fixed to the average value $\langle n \rangle=1$. 
As known, the ABP dynamics in polar coordinates is simply a diffusive process for $\theta$:
\begin{equation}
\dot{\theta} = \sqrt{\frac{2}{\tau}}\, \xi \,,
\label{eq_thetaABP}
\end{equation}
where $\xi$ is a white noise with unit variance and zero average and the time scale $\tau=1/D_\text{r}$ represents the persistence time induced by the rotational diffusion coefficient $D_\text{r}$.

\paragraph*{Active Ornstein-Uhlenbeck Particles (AOUPs).} In the case of AOUPs, $\mathbf{n}$ is represented by a two-dimensional Ornstein-Uhlenbeck process that allows both the modulus $n$ and the orientation $\theta$ to fluctuate with related amplitudes. 
The AOUP distribution is a two-dimensional Gaussian such that each component fluctuates around a vanishing mean value with unitary variance. In polar coordinates, the
probability distribution of the AOUP self-propulsion reads: 
\begin{equation}
P_\text{AOUP}(n, \theta) \sim \frac{1}{2\pi} n \exp{\left(-n^2 \right)} \,.
\end{equation}
The dynamics $\dot{\mathbf{n}}= - \frac{\mathbf{n}}{\tau} + \sqrt{\frac{1}{\tau}} \boldsymbol{\chi}$ generating the process is usually written in Cartesian coordinates, where $\boldsymbol{\chi}$ is a two-dimensional vector of white noises with uncorrelated components having unitary variance and zero average.
To shed light on the relation with the ABP, it is convenient to express the dynamics of AOUP in polar coordinates (It\^{o} integration):
%\begin{equation}
\begin{subequations}
\begin{align}
\dot{n}&= - \frac{n}{\tau} + \sqrt{\frac{1}{\tau}} \chi_n + \frac{1}{2\tau n} \\
\dot{\theta}&=\sqrt{\frac{1}{\tau}} \frac{\chi_\theta}{n} \,,
\label{eq_AOUPpolar}
\end{align} 
\label{eq_AOUPpolarALL}
\end{subequations}
%\end{equation}
where $\chi_n$ and $\chi_\theta$ are white noises with unit variance and zero average.
While still being coupled to the dynamics of $n$,
the angular equation for $\theta$ is quite similar to that describing the ABP dynamics in Eq.~\eqref{eq_thetaABP}.

\paragraph*{Mapping between ABPs and AOUPs.} 
Usually, the connection between ABPs and AOUPs is established by demanding that the steady-state temporal correlations of the self-propulsion velocity $v_0\mathbf{n}$ of ABPs and AOUPs are equal. 
Note that, by introducing this generic form of the active force in Eq.~\eqref{eq_t1}, we have already included in the dynamics the mapping $2D_a/\tau\! =\! v_0^2$ through which we have eliminated the active diffusivity $D_a$ from the conventional notation for the AOUP dynamics. 
Likewise, the second relation $D_r = 1/\tau$ is implied in Eq.~\eqref{eq_thetaABP}.
As a result, both models share the same autocorrelation function
\begin{align}
\langle \mathbf{n}(t)\cdot\mathbf{n}(0)\rangle=e^{-\frac{t}{\tau}}
\label{eq_nncorr}
\end{align} 
of the self-propulsion vector $\mathbf{n}$, despite possessing different distribution $P_\text{ABP}(n, \theta)\neq P_\text{AOUP}(n, \theta)$.
Apart from this mapping, there is currently no apparent deeper relation between the stochastic processes,
Eq.~\eqref{eq_thetaABP} and Eq.~\eqref{eq_AOUPpolar}, underlying the dynamics of ABP and AOUP, respectively. As a next step, we establish such a connection by introducing a more general model.

\paragraph*{Unification in the parental active model (PAM).}

\begin{figure}[t!]
\includegraphics[width=\columnwidth]{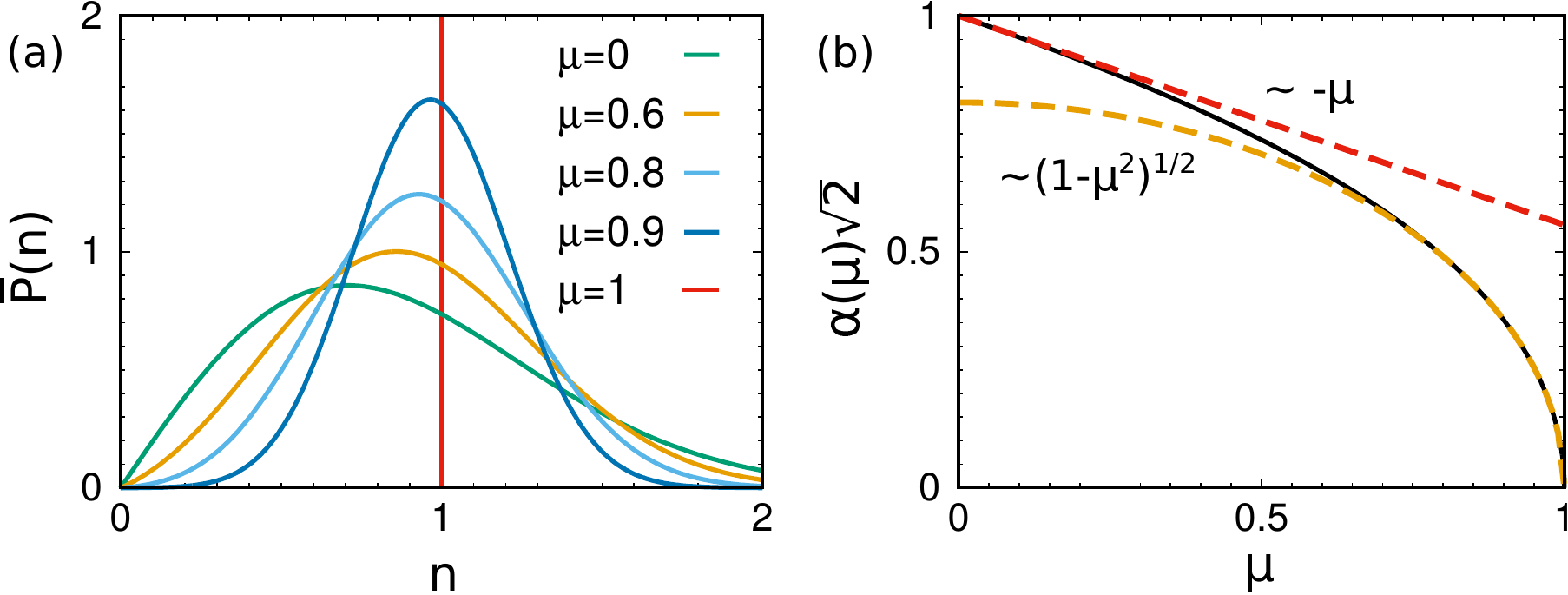}
\caption{
Stationary solution for the self-propulsion vector $\mathbf{n}$ in the PAM.
Panel~(a): distribution $\bar{P}(n)=\int_0^{2\pi} d\theta P(n,\theta)$,  
given by  Eq.~\eqref{eq_PAM}, of the radial component $n=|\mathbf{n}|$ 
for different values of $\mu$, interpolating between  AOUP  ($\mu=0$) and ABP  ($\mu=1$). 
Panel~(b): relation between the parameters $\alpha$ and $\mu$, which guarantees that $\langle n^2 \rangle =1$, leaving the velocity scale $v_0$ invariant. Red and yellow dashed curves indicate the asymptotic solutions for $\mu \to 0$ and $\mu \to 1$, respectively, given by Eq.~\eqref{eq:alphamu_approx}. }
\label{fig:PMdist}
\end{figure}

Now, we are ready to define a ``parental'' active model (PAM) from which one can recover both ABPs and AOUPs, as limiting cases. 
The most natural steady-state distribution for a PAM accounting for these features simply introduces Gaussian fluctuations and reads:
\begin{equation}
P(n, \theta) \sim \frac{n}{2\pi} \exp{\left(-\frac{\left(n-\mu\right)^2}{2\alpha^2}  \right)} \,.
\label{eq_PAM}
\end{equation}
 This is one of the most simple distributions that allow the modulus to fluctuate around a nonzero attraction point, $\mu$, with modulus fluctuations, $\alpha^2$, which are independent of those of the active force direction $\theta$. Note that $P(n, \theta)$ is constant in $\theta$ so that $P(n, \theta)\!\sim\! \bar{P}(n)$, where $\bar{P}\!=\!\int_0^{2\pi} d\theta P$ is the reduced distribution of the self-propulsion velocity modulus, cf.~Fig.~\ref{fig:PMdist}.

The dynamics of the PAM, i.e., the dynamics which generate the steady-state distribution~\eqref{eq_PAM} in polar coordinates are (It\^{o} integration):
\begin{subequations}
\begin{align}\label{eq_PAMn}
\dot{n}&= - \frac{\left(n-\mu\right)}{\tau} + \sqrt{\frac{2\alpha^2}{\tau}} \chi_n + \frac{\alpha^2}{\tau n}\,, \\
\dot{\theta}&=\sqrt{\frac{2f(\alpha)}{\tau}} \frac{\chi_\theta}{n}  \,,
\label{eq_PAMtheta}
\end{align}
\label{eq_PAMall}
\end{subequations}
where $f(\alpha)=1-\alpha^2$ and $\alpha\in[0,1/\sqrt{2}]$.
 The representation of Eq.~\eqref{eq_PAMall} in Cartesian coordinates is discussed in Appendix~\ref{app_PAMC}.
We remark that the shape of $f(\alpha)$ cannot be arbitrary, but should guarantee that the total noise strength remains constant throughout all offspring of the PAM, namely $\alpha^2+f(\alpha)=1$.
Fixing $\alpha=1$ and $\mu=0$ the dynamics coincides with that of the AOUP, cf.~Eq.~\eqref{eq_AOUPpolarALL}.
For $\alpha=0$ and $\mu=1$, we obtain the ABP dynamics, because the time evolution, Eq.~\eqref{eq_PAMn}, of $n$ has the solution
%\begin{equation}
$n(t)=1+\exp{\left( -t/\tau  \right)}$, %\,,
%\end{equation}
so that $n$ approaches deterministically the ABP unit value.
Then, the dynamics, Eq.~\eqref{eq_PAMtheta}, for the angle $\theta$ reverts to Eq.~\eqref{eq_thetaABP}.

To identify the number of free parameters in our general PAM, we demand that the typical speed $v_0$ induced by the self-propulsion remains as a fixed velocity scale. To achieve this, we relate the modulus fluctuations $\alpha$ to the peak position $\mu$ by requiring $\langle n^2 \rangle =1$ (otherwise $v_0$ would have to be renormalized). The resulting relation $\alpha(\mu)$ (see Appendix~\ref{app_amu}), leaves $\mu$ as the only free parameter of the PAM.
Near the two limiting cases of the AOUP ($\mu \to 0$) and ABP ($\mu \to 1$), the relation $\alpha(\mu)$ simplifies and reads:
\begin{equation}
\label{eq:alphamu_approx}
\alpha \approx  
\begin{cases}
\frac{1}{\sqrt{2}} \left( 1- \frac{\sqrt{\pi}}{4} \mu \right), & \mu \to 0, \\
\sqrt{\frac{1 - \mu^2}{3} },  & \mu \to 1 \,.
\end{cases}
\end{equation}
 In Fig.~\ref{fig:PMdist}~(b), we compare these simple representations to $\alpha(\mu)$, obtained by solving numerically $\langle n^2 \rangle =1$, and we find good agreement for $0\leq\mu\lesssim0.3$ and $0.7\lesssim\mu\leq1$.
The resulting steady-state distributions are shown in Fig.~\ref{fig:PMdist}~(a) for different $\mu$, interpolating between AOUPs (green curve) and ABPs (red curve), see also Fig.~\ref{fig:picture} for the representation in Cartesian coordinates.

Apart from the free parameter $\mu$,
which uniquely characterizes each descendant of the PAM,
the whole family of models shares
a common persistence time $\tau$ of the active motion and an equal dynamical correlation, given by Eq.~\eqref{eq_nncorr}.
As a result, some basic dynamical properties for a potential-free particle are the same for each value of $\mu$, such as the velocity autocorrelation function, the mean and the mean-squared displacements, in accordance with the well-known results in the limiting cases of ABPs~\cite{ten2011brownian, sevilla2015smoluchowski} and AOUPs~\cite{fodor2018statistical}.

\begin{figure}[t!]
\includegraphics[width=\columnwidth]{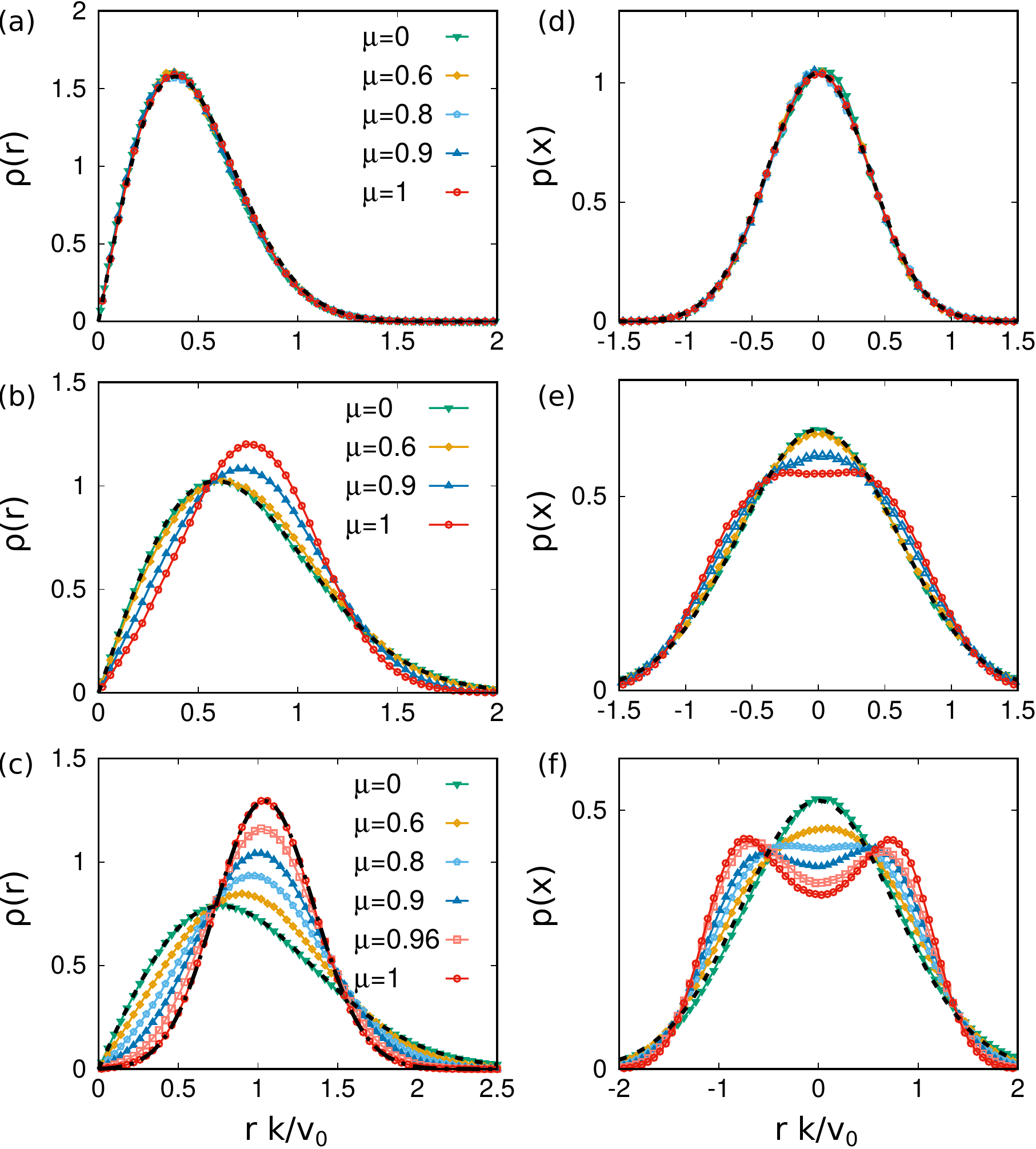}
\caption{
Probability distribution of the active particle position in a harmonic external potential. Panel~(a),~(b) and~(c) show the radial density distribution $\rho(r)$, as a function of $r k/v_0$, while panels~(d),~(e) and~(f) plot the distribution (projected onto one axis) $p(x)$ as a function of $x k/v_0$. Panels~(a) and~(d) are obtained with $k\tau=10^{-1}$, panels~(b) and~(e) with $k\tau=1$ and, finally, panels~(c) and~(f) with $k\tau=10^2$. The black dashed lines in all the panels are obtained by Eq.~\eqref{eq:AOUP_harmonic_largepersistence}, while the black dashed-dotted line in panel~(c) by Eq.~\eqref{eq:ABP_harmonic_largepersistence}. 
Panels (a) and~(d), (b) and~(e), (c) and~(f) share the same legend.
}
\label{fig:varyingmu}
\end{figure}

\paragraph*{PAM in harmonic confinement.}
The main difference between ABPs and AOUPs occurs in the dynamics of the radial component of the active force. The consequences of that become highly relevant if the particle is subject to an additional, external potential. 
As a reference study, we confine the system via a harmonic trap, so that the external force $\mathbf{F}(\mathbf{x})=-k \,\mathbf{x}$ is exerted on the active particle. 
 The curvature of the potential $k$ introduces an additional time scale that is recast onto a dimensionless parameter $k\tau$ controlling the dynamics.  
In Fig.~\ref{fig:varyingmu}, we study the radial probability distribution, $\rho(r)$, and the reduced distribution in Cartesian coordinates, $p(x)$, projected onto the $x$-axis, 
for different values of $\mu$ and $k\tau $.

Before discussing the behavior of the generic PAM in detail, we provide further analytic insight on the extreme cases (calculations are reported in Appendices~\ref{app_AOUP} and~\ref{app_ABP}).
As a Gaussian process, the AOUP gives rise to the exact solution \cite{das2018confined, caprini2019comparative, dabelow2021irreversible}:
\begin{equation}
\label{eq:AOUP_harmonic_largepersistence}
    \rho(r)\sim \exp{\left(-\frac{k \Gamma}{\left( D \Gamma + \frac{v_0^2\tau}{2} \right)}\frac{r^2}{2}\right)} \,,
\end{equation}
where as usual in AOUP systems $\Gamma=1+k\tau $, plays the role of an effective friction coefficient~\cite{caprini2019activity}.
Assuming large persistence, $k\tau  \gg 1$, we further develop an analytical prediction for the ABP: 
\begin{equation}
\label{eq:ABP_harmonic_largepersistence}
    \rho(r)\sim r^{1/2} \exp{\left(-\left(k+\frac{1}{2\tau}\right) \frac{1}{2D}\left(r-\frac{v_0}{k+\frac{1}{2\tau}}\right)^2\right)} \,
\end{equation}
which reflects the bimodality of the density distribution~\cite{malakar2020steady, pototsky2012active, hennes2014self, hennes2014self, rana2019tuning, basu2019long, santra2021direction} (see also Refs.~\cite{takatori2016acoustic, dauchot2019dynamics} for experimental studies) as a distinct feature compared to the Gaussian shape of the AOUP solution.

When the active force relaxes faster than the particle position, such that $k\tau  \ll 1$, the dynamical details of the active force in the generic PAM cannot affect the distribution, which is thus independent of $\mu$, as shown in Fig.~\ref{fig:varyingmu}~(a) and~(d).
In this regime, the shape of $\rho(r)$ (or equivalently $p(x)$) coincides with the analytical AOUP result, Eq.~\eqref{eq:AOUP_harmonic_largepersistence} with $\Gamma \to 1$, for every $\mu$. This approximation can be explicitly derived also in the opposite extreme case of ABPs (see Appendix~\ref{app_ABP}).
This occurs because the active force behaves as a noise term and thus it only modifies the variance of $\rho(r)$ with respect to the passive case, in the spirit of an effective temperature.
In the intermediate persistence regime, $k\tau \!\sim\! 1$, Fig.~\ref{fig:varyingmu}~(b) and Fig.~\ref{fig:varyingmu}~(e) indicate that the density gradually departs from its Gaussian form, given by Eq.~\eqref{eq:AOUP_harmonic_largepersistence}, when $\mu$ is increased:
the position of the main peak of $\rho(r)$ shifts towards larger values of $r$ while the shape $p(x)$ displays the onset of bimodality. 
These differences become most significant in the large persistence regime, $k\tau  \gg 1$, where the ABP solution is well-represented by Eq.~\eqref{eq:ABP_harmonic_largepersistence}, roughly centered around $v_0/[k+1/(2\tau)] \to v_0/k$ (for $k\tau  \gg 1$).
Also for smaller $\mu$, the radial density $\rho(r)$ has a strongly non-Gaussian shape, see Fig.~\ref{fig:varyingmu}~(c). 
We further show in Fig.~\ref{fig:varyingmu}~(f) that for a large persistence, even a small increase of $\mu$ induces drastic changes in the shape of $p(x)$, eventually inducing a unimodal $\to$ bimodal transition.

\begin{figure}[t!]
\includegraphics[width=\columnwidth]{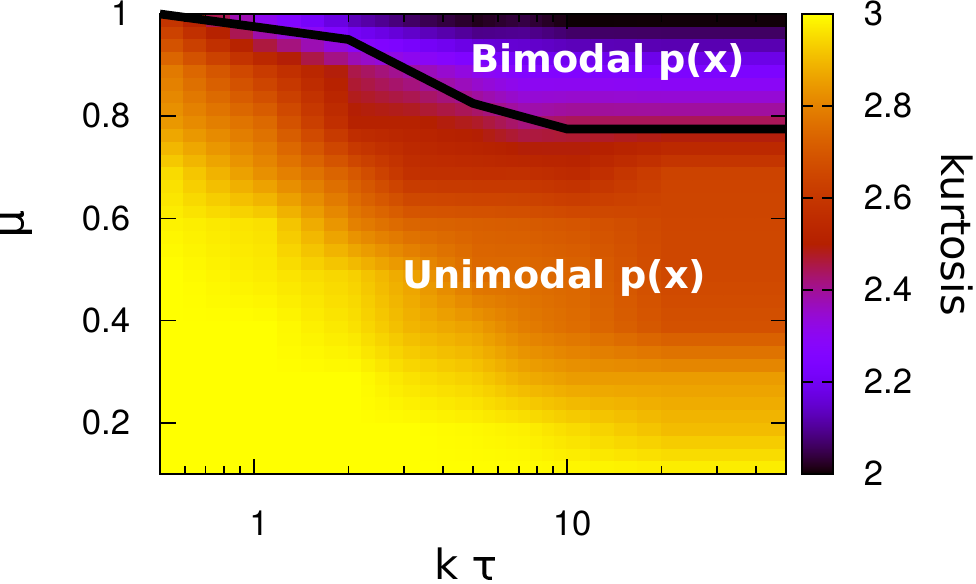}
\caption{
Two-states phase diagram of the active harmonic oscillator, by varying $k\tau$ and $\mu$ (and, thus, $\alpha(\mu)$ accordingly), distinguishing between regions where the spatial distribution, $p(x)$, is unimodal and bimodal, as explicitly indicated in the graph.
The two regions are separated by a solid black line, $\mu_c(\tau_c)$, tracked in correspondence of the first value of $k\tau$ such that $p(x)$ shows a bimodality: in practice, we fit the exponential of a fourth order polynomial $\exp(-a x^4 + b x^2 + c)$, identifying a point on the critical line $\mu_c(\tau_c)$ as the smaller value of $\mu$ (for each $k\tau$) such that $b<0$. 
In addition, we plot the kurtosis of $p(x)$, namely $\langle x^4 \rangle/\langle x^2 \rangle^2$, as a color gradient.
We remark that the typical values of the kurtosis in correspondence of the transition line are between 2.3 and 2.5. 
}
\label{fig:PDink}
\end{figure}

 In Fig.~\ref{fig:PDink}, such a transition is depicted through a phase diagram as a function of $\mu$ and $k\tau$, distinguishing between unimodal and bimodal configurations and showing the kurtosis of $p(x)$ as a color gradient.
For small values of $\mu$, the distribution $p(x)$ is unimodal (region $1$) independently of $k\tau$. Starting from $\mu=0$ (AOUP model) which is Gaussian, the increase of $\mu$ induces non-Gaussianity in the shape of $p(x)$, which reflects onto the decrease of the kurtosis to values smaller than 3.
However, while for small values of $k\tau$, $p(x)$ still remains unimodal upon increasing $\mu$ (compare Fig.~\ref{fig:varyingmu}~(d)),  a transition towards a bimodal distribution, which is characterized by kurtosis values $\sim 2$, takes place (region 2) as soon as $k\tau \sim 1$.
The corresponding critical curve $\mu_c(\tau_c)$ (black line in Fig.~\ref{fig:PDink}) 
decreases when $k\tau$ is increased until reaching a plateau for $k\tau \gg 1$. This is consistent with Eqs.~\eqref{eq:AOUP_harmonic_largepersistence} and~\eqref{eq:ABP_harmonic_largepersistence} which both do not depend on $k\tau$ for $k\tau  \gg 1$.
In general, the fluctuation of the modulus $n$ of the self-propulsion vector inhibits the ability of the active particle to stay far from the potential minimum, even in the harmonic oscillator case.

\paragraph*{Conclusions.}
We developed a unifying parental active model (PAM)
for the stochastic dynamics of active particles.
This PAM shows that the established ABPs and AOUPs descriptions stand on an equal level 
as being sisters rather than cousins.
The family of explored models shares defining properties of active matter, such as the exponential dynamical correlations on the scale of the persistence time $\tau$ and the common velocity scale $v_0$.
The only differences lie in the modulus distributions of the self-propulsion velocity, which can be continuously transferred from a Gaussian form (AOUP) to a sharp peak (ABP) by sweeping a single parameter.
As a benchmark study, we examined the stationary distribution in a harmonic potential and mapped out the
transition between unimodal and bimodal, which marks the classical "failure" of AOUPs to reproduce the behavior of ABPs in the large-persistence regime.

For the purpose of realistic modeling, however, both AOUPs and ABPs are idealized. 
This is because a perfectly constant modulus of the self-propulsion velocity is highly unlikely due to the individual nature of biological agents and various types of fluctuations.
Bacteria, for example, can display fairly broad~\cite{wu2009periodic, theves2013bacterial} or even bimodal~\cite{ipina2019bacteria, otte2021statistics} speed distributions.
Also macroscopic agents like locusts~\cite{bazazi2011}, whirligig beetles~\cite{devereux2021} or zebrafish~\cite{zampetaki2021collective,mwaffo2017,burbano2020} exhibit natural speed fluctuations.
To realistically describe these systems, a theoretical approach should incorporate both fluctuations of the modulus and the direction of the self-propulsion velocity~\cite{romanczuk2011,Shee2021self,breoni2020active,mwaffo2017,burbano2020}.
For this purpose, our description within the PAM is particularly convenient, 
because it is based on a single stochastic process $\mathbf{n}$ of unit standard deviation (i.e., $v_0$ is treated as a velocity scale and does not fluctuate itself), such that all descendant models with an intermediate value of the parameter $\mu$
can be evaluated with the same numerical effort as ABPs and AOUPs.

The family of models can be systematically extended by realizing that the PAM merely gives rise to more diversity in the stationary properties of the underlying stochastic process, while the autocorrelation~\eqref{eq_nncorr} of the self-propulsion velocity remains equal for all offspring.
Another common model of active particles involves the run and tumble motion~\cite{tailleur2008statistical,solon2015,angelani2017confined, gradenigo2019first} where the autocorrelation is a step function because, after running for a straight path, the particle instantaneously changes the direction of its active velocity after a typical tumbling rate.
In our line of reasoning, this particular shape (at the same persistence time scale $\tau$, related to the inverse of the tumbling rate) of the dynamical autocorrelation function could be viewed as, say, a different gender.
In practice, the notion of run-and-tumble-like dynamics can be easily combined with our PAM by drawing after each tumbling event the new direction and modulus of the self propulsion vector according to the stationary distribution in Eq.~\eqref{eq_PAM}.

In conclusion, the PAM both unifies ABPs and AOUPs and provides a crucial step towards more realistic modeling of overdamped (dry) active motion in general, which should in future work be employed to provide an improved fit of experimental swim-velocity distributions. 
 Investigating the effect of the swim-velocity fluctuations could represent an interesting perspective for circle swimming~\cite{kummel2013circular, lowen2016chirality, banerjee2017odd, kurzthaler2017,liao2018clustering, reichhardt2019reversibility}, systems with spatial-dependent swim velocity~\cite{lozano2016phototaxis, stenhammar2016light,sharma2017brownian, vizsnyiczai2017light,soker2021activity,caprini2021dynamics}, and inertial dynamics~\cite{takatori2017inertial, lowen2020inertial, gutierrez2020inertial, dai2020phase, su2020inertial} even affecting the orientational degrees of freedom~\cite{scholz2018inertial, sprenger2021time}. 
The generalization of PAM to these cases could be responsible for new intriguing phenomena which will be investigated in future works.

\acknowledgments
We thank Alexandra Zampetaki for helpful discussions.
LC acknowledges support from the Alexander Von Humboldt foundation.
HL and RW acknowledge support by the Deutsche Forschungsgemeinschaft  (DFG)  through  the  SPP  2265,  under  grant  numbers LO 418/25-1 (HL)  and  WI5527/1-1 (RW).

\appendix

\section{PAM dynamics in Cartesian coordinates \label{app_PAMC}}
In this appendix, we report the expression for the PAM dynamics in Cartesian coordinates. Applying Ito calculus, we obtain:
\begin{equation}
\label{eq:parentaldynamics_cartesian}
\begin{aligned}
\dot{\mathbf{n}}=& - \frac{1}{\tau} \left( \mathbf{n} - \mu \frac{\mathbf{n}}{n} \right) + \frac{\mathbf{n}}{n^2}\frac{1}{\tau} \left( \alpha^2-f(\alpha) \right) \\
&+\sqrt{\frac{2 \alpha^2}{\tau}}
\frac{\mathbf{n}}{n} \chi_n + \sqrt{\frac{2 f(\alpha)}{\tau}} \boldsymbol{R} \cdot \frac{\mathbf{n}}{n} \chi_\theta
\end{aligned}
\end{equation}
being $\boldsymbol{R}$ the rotational matrix of 90 degree.
Alternatively, the last term can be expressed in a more familiar form in terms of the cross product:
$$
\boldsymbol{R} \cdot\frac{\mathbf{n}}{n} \chi_\theta = \frac{\mathbf{n}}{n} \times \hat{z} \chi_\theta \,.
$$
By setting $\mu=0$ and $\alpha=1/\sqrt{2}$ in Eq.~\eqref{eq:parentaldynamics_cartesian}, we recover the AOUP model. Indeed, only the term $-\mathbf{n}/\tau$ survives on the first line while the noise terms in the second line reduces to a vector of white noise because any orthogonal transformation applied on a vector of white noises is still a vector of white noise.
Instead, by setting $\mu=1$ and $\alpha=0$ in Eq.~\eqref{eq:parentaldynamics_cartesian}, only the term $-\mathbf{n}/\tau$ survives on the first line, because $n^2=n=1$ and only the second noise survives on the second line, so that we obtain the ABP equation  (It\^{o} integration)  
\begin{equation}
\dot{\mathbf{n}}= - D_r \mathbf{n}+\sqrt{2D_r} \mathbf{n} \times \mathbf{z}\, \xi
\end{equation}
in Cartesian coordinates, where $\mathbf{z}=(0,0,1)$.

\section{Obeying the unit-variance condition 
\label{app_amu}}
In this appendix, we give the analytic expression of the second moment $\langle n^2 \rangle$ of the PAM distribution (see Eq.~\eqref{eq_PAM}) needed to impose the constraint $\langle n^2\rangle=1$ dictated by the given velocity scale $v_0$.
After algebraic manipulations, we get:
\begin{equation}
	\langle n^2 \rangle =  3 \alpha^2 + \mu^2 - \mathcal{N} \alpha^4 e^{- \frac{\mu^2}{2 \alpha^2} }\,,
\end{equation}
where $\mathcal{N}$ is the normalization constant of the distribution~\eqref{eq_PAM}, which explicitly reads: 
\begin{equation}
	\mathcal{N}^{-1} =  \frac{\alpha^2}{2} \frac{\mu}{\sqrt{2}\alpha} \left( 4 \sqrt{\pi} + \Gamma\left(-\frac{1}{2}, \frac{\mu^2}{2 \alpha^2} \right) \right)\,.
\end{equation}
Here, $\Gamma\left(s,x\right)$ denotes the upper incomplete gamma function. 
The condition requiring $\langle n^2 \rangle =1$ follows as 
\begin{equation}
3 \alpha^2 + \mu^2 - \mathcal{N} \alpha^4 e^{- \frac{\mu^2}{2 \alpha^2} } = 1,
\end{equation}
which is solved for $\alpha(\mu)$ in Fig.~\ref{fig:PMdist} 
and yields the asymptotic solutions near the two limiting cases of the AOUP ($\mu \to 0$) and ABP ($\mu \to 1$) model given by Eq.~\eqref{eq:alphamu_approx}.

\section{AOUP in a harmonic potential \label{app_AOUP}}

Here, we provide the solution of Eq.~\eqref{eq_t1} with the external force $\mathbf{F}(\mathbf{x})=-k \,\mathbf{x}$.
In the AOUP case (or the PAM with $\mu=0$ and thus $\alpha=1/\sqrt{2}$), the dynamics can be solved exactly, because of its linearity.
The whole solution for the probability distribution $\mathcal{P}(\mathbf{x}, \mathbf{n})$ reads:
\begin{equation}
\begin{aligned}
    \mathcal{P}(\mathbf{x}, \mathbf{n})=&\mathcal{N} \exp{\left(-\frac{\Gamma k}{\Gamma D +\frac{v_0^2\tau}{2}} \frac{r^2}{2}\right)}\\
    &\times\exp{\left(-\frac{\Gamma}{v_0^2} \left(\mathbf{n}- \frac{k}{2}\frac{\Gamma v_0^2 \tau}{\left(\frac{v_0^2\tau}{2}+D\right)}\mathbf{x}  \right)\right)},
\end{aligned}
\end{equation}
where $r^2=x^2+y^2$ in two spatial dimensions.
By integrating out the self-propulsion vector $\mathbf{n}$ and switching to polar coordinates, 
we easily obtain the expression for the radial probability distribution, $\rho(r)$, which reads:
\begin{equation}
\label{app:rho(r)AOUP}
\rho(r)=\mathcal{N} \exp{\left(-\frac{k \Gamma}{\left( D \Gamma + \frac{v_0^2\tau}{2} \right)}\frac{r^2}{2}\right)}
\end{equation}
where $\Gamma$ plays the role of an effective friction coefficient and reads:
\begin{equation}\label{app:rho(r)AOUP_Gamma}
    \Gamma=1+k \tau\,,
\end{equation}
as stated in Eq.~\eqref{eq:AOUP_harmonic_largepersistence} of the main text.
From Eq.~\eqref{app:rho(r)AOUP}, we can identify an effective temperature, say the variance of the distribution, as
\begin{equation}
T_{\text{eff}} =  \left( D + \frac{v_0^2\tau}{2 \Gamma}\right)\,.
\end{equation}

\section{ABP in a harmonic potential \label{app_ABP}}

To get analytical results in the case of an ABP (or the PAM with $\mu=1$ and thus $\alpha=0$) in a harmonic trap, it is convenient to express the positional dynamics \eqref{eq_t1} in polar coordinates, $(x,y) \to (r, \phi)$, such that $r=\sqrt{x^2+y^2}$ and $\phi=\text{atan}{\frac{y}{x}}$. 
Applying Ito calculus to the dynamics~(1) of the main text to perform the change of variables, one gets: 
\begin{subequations}
\label{eq:polarcomponents_position}
\begin{align}
&\dot{r}= - k\,r + \frac{D}{r} + v_0 \cos{(\theta-\phi)} + \sqrt{2D} w_r\\
&\dot{\phi}=v_0 \frac{\sin{(\theta-\phi)}}{r} + \frac{\sqrt{2D}}{r} \sqrt{2D} w_\phi \,,
\end{align}
\end{subequations}
where the orientation $\theta$ of the (normalized) self-propulsion vector $\mathbf{n}$ evolves according to Eq.~\eqref{eq_thetaABP}.
From here, the Fokker-Planck equation for the probability distribution, $p=p(r,\phi, \theta)$, reads:
\begin{equation}
\label{app:FP_radial_harmonic}
\begin{aligned}
\partial_t p&= \partial_r \left[k\,r - \frac{D}{r} - v_0 \cos{(\theta-\phi)} + D\partial_r   \right] p\\
&+\partial_\phi \left[ \frac{D}{r^2}\partial_\phi - \frac{v_0}{r} \sin{(\theta-\phi)} \right] p + \frac{1}{\tau} \partial^2_\theta p \,.
\end{aligned}    
\end{equation}

Separating angular and radial currents in Eq.~\eqref{app:FP_radial_harmonic} allows us to find approximated solutions for the conditional angular probability distribution $f(\theta-\phi|r)$ (i.e., the angular probability distribution at fixed radial position $r$),
which we will  use later to estimate the radial density distribution $\rho(r)$.
In other words, by setting the second line in Eq.~\eqref{app:FP_radial_harmonic} equal to zero, we obtain:
\begin{equation}
    \label{eq:anuglar_distribution}
    f(\theta-\phi|r)=\mathcal{N} e^{ a \cos{(\theta-\phi)} }
\end{equation}
where $a$ reads:
\begin{equation}
    a=\frac{v_0 }{D}\frac{r}{\left( 1+\frac{r^2}{D\tau} \right)}\,.
\end{equation}
In the small persistence regime, $k\tau \ll1$, this distribution converges to a flat profile, because $a\to 0$ vanishes. This reflects the fact that both $\theta$ and $\phi$ are uniformly distributed and, thus, also their difference. 
Instead, in the large persistence regime, Eq.~\eqref{eq:anuglar_distribution} is peaked around $\phi \sim \theta$ and its variance becomes smaller as $k\tau $ is increased.

As a first step to finding an approximation for $\rho(r)$,
we now calculate the average
\begin{equation}
    \langle \cos(\theta-\phi)\rangle = \frac{I_1(a)}{I_0(a)} 
    \label{eq_costhph}
\end{equation}
with respect to the conditional angular distribution, Eq.~\eqref{eq:anuglar_distribution}, where $I_0(a)$ and $I_1(a)$ are the modified Bessel function of the first kind of order 0 and 1, respectively.
With this result, we can achieve the derivation starting directly from Eq.~\eqref{app:FP_radial_harmonic}. 
At first, we assume the zero current condition along for the radial current, namely we set to zero the first line in Eq.~\eqref{app:FP_radial_harmonic}. Then, we replace $\cos(\theta-\phi)\to \langle \cos(\theta-\phi) \rangle$, where we approximate the result from Eq.~\eqref{eq_costhph} in two different regimes.

\subsection{Small-persistence regime}

In the small persistence regime, such that $k\tau  \ll1$, we have $a\ll1$ and we can approximate:
\begin{equation}
\label{app:avcos_smalltau}
\langle \cos(\theta-\phi)\rangle=\frac{I_1(a)}{I_0(a)}\approx \frac{a}{2}=\frac{1}{2}\frac{v_0}{D}\frac{r}{\left(1+\frac{r^2}{D\tau} \right)} \,.
\end{equation}
The small persistence time regime further allows us to replace $r^2 \to \langle r^2\rangle=\frac{1}{k}\left(D+ \frac{v_0^2 \tau}{2} \right)$ in Eq.~\eqref{app:avcos_smalltau}.
The expression for $\langle r^2\rangle$ is achieved by recalling that the active particle in the small persistence regime is subject to the effective temperature $D+v_0^2\tau/2$, a result holding for a general potential.
From here, the zero current condition in Eq.~\eqref{app:FP_radial_harmonic} leads to an equation for $\rho(r)$:
\begin{equation}
    \left[  k^* r  - \frac{D}{r} + D\frac{\partial}{\partial r}  \right] \rho(r)=0 \,,
\end{equation}
where:
\begin{equation}
    k^*=k+\frac{v_0^2}{D+\frac{1}{k\tau}\left(D+\frac{v_0^2\tau}{2}  \right)}     \,.
\end{equation}
This equation can be easily solved obtaining an expression for $\rho(r)$ that after algebraic manipulation reads:
\begin{equation}
\label{app:rho_smalltau}
\rho(r)=\mathcal{N} \exp{\left(-\frac{k \Gamma}{\left( D \Gamma + \frac{v_0^2\tau}{2} \right)}\frac{r^2}{2}\right)},
\end{equation}
where $\Gamma=1+k \tau \to 1$ is defined according to Eq.~\eqref{app:rho(r)AOUP_Gamma}.
This distribution coincides with the AOUP one~\eqref{app:rho(r)AOUP}.
%$$
%\Gamma=1+k \tau \to 1\,.
%$$

We observe that in the limit of very small $\tau$, the above result~\eqref{app:rho_smalltau} coincides with that obtained in the passive limit, which can be achieved by setting $v_0 \to 0$.
In this case, we have $a \to 0$ and thus $\langle \cos(\theta-\phi) \rangle=0$ in Eq.~\eqref{app:FP_radial_harmonic} (and the same for the sinus) because $\theta$ is uniformly distributed between $0$ and $2\pi$.
Therefore, Eq.~\eqref{eq:polarcomponents_position} simply converges onto the equation of a passive particle holding for $v_0^2\tau \ll D$.
We further remark that our result is consistent with that obtained by the hydrodynamic approach holding in the case of ABP in the regime of small $\tau$, which allows us to recover Eq.~\eqref{app:rho_smalltau} with $\Gamma \to 1$.

\subsection{Large-persistence regime}

In the large persistence case, $k\tau \gg 1$, the self-propulsion relaxes much slower than the position distribution.
Also in this case, we can adopt the same strategy used in the small persistence regime with the crucial difference that now we have $a\gg 1$, so that we can approximate Eq.~\eqref{eq_costhph} as:
\begin{equation}
 \langle \cos(\theta-\phi)\rangle = \frac{I_1(a)}{I_0(a)} \approx 1-\frac{1}{2a} = 1 - \frac{1}{2} \frac{r}{v_0} \left( \frac{D}{r^2}+\frac{1}{\tau}\right) .   
\end{equation}
Plugging this result into Eq.~\eqref{app:FP_radial_harmonic} and using the zero-current condition allows us to find the equation for the radial density, $\rho(r)$, which reads:
\begin{equation}
    \left[  r \left(k+\frac{1}{2\tau}\right) - \frac{D}{r^{1/2}} -v_0 + D\frac{\partial}{\partial r}  \right] \rho(r)=0 \,
\end{equation}
and whose solution can be explicitly obtained:
\begin{equation}
\label{app:radialprob_largetau}
    \rho(r)=\mathcal{N} r^{1/2} \exp{\left(-\left(k+\frac{1}{2\tau}\right) \frac{1}{2D}\left(r-\frac{v_0}{k+\frac{1}{2\tau}}\right)^2\right)} \,.
\end{equation}
Here, the result is fairly different from the Gaussian distribution~\eqref{app:rho(r)AOUP} obtained in the case of AOUP dynamics.
The profile~\eqref{app:radialprob_largetau} is well-approximated by a Gaussian centered at $r=v_0/(k+1/2\tau)$ with variance $D/(k+1/2\tau)$.

Note that the result~\eqref{app:radialprob_largetau} is almost consistent with that obtained in Ref.~\onlinecite{caprini2019comparative} in the limit $\tau \to \infty$. 
However, with respect to Ref.~\onlinecite{caprini2019comparative}, here we improve the approximation for the angular distribution that leads to  a prefactor $r^{1/2}$ (instead of simply $r$) which is in better agreement with data. 
To establish a closer relation to this result, we remark that, in the large persistence regime, the  angular distribution~\eqref{eq:anuglar_distribution} derived here  
can be further approximated by a Gaussian by expanding the cosine around $\theta\sim\phi$:
\begin{equation}
f(\theta-\phi|r)=\mathcal{N} e^{-\frac{a}{2}(\theta-\phi)^2} \,.
\end{equation}
The expression for $\rho(r)$ resulting from this approximation is then consistent with the previous prediction~\cite{caprini2019comparative}
in the large persistence regime.

\bibliography{SD.bib}

\end{document}